# USING A NAMESERVER TO ENHANCE CONTROL SYSTEM EFFICIENCY*

J. Sage, M. H. Bickley, K. S. White, Jefferson Lab, Newport News, VA 23606, USA

Abstract

The Thomas Jefferson National Accelerator Facility (Jefferson Lab) control system uses a nameserver to reduce system response time and to minimize the impact of client name resolution on front-end computers. The control system is based on the Experimental Physics and Industrial Control System (EPICS), which uses name-based broadcasts to initiate data communication. By default, when EPICS process variables (PV) are requested by client applications, all front-end computers receive the broadcasts and perform name resolution processing against local channel name lists. The nameserver is used to offload the name resolution task to a single node. This processing, formerly done on all front-end computers, is now done only by the nameserver. In a control system with heavily loaded front-end computers and high peak client connection loads, a significant performance improvement is seen. This paper describes the name server in more detail, and discusses the strengths and weaknesses of making name resolution a centralized service.

## 1 INTRODUCTION

The Jefferson Lab Accelerator, Cryogenics Plant, Free Electron Laser and Experimental Halls are controlled by over two hundred computers in a distributed system based on EPICS[1]. These computers either serve as back-end computers for operational consoles, running system and high-level functions, or front-end computers for device control[2]. In our EPICS based control system, the front-end computers are referred to as Input/Output Controllers (IOCs). Due to heavy and ever increasing CPU loads on many IOCs, we developed software to remove the computation load associated with process variable name resolution from these front-end computers.

## 2 BACKGROUND

The Jefferson Lab Accelerator Control System uses approximately 100 front-end computers for device control and algorithm execution. These IOCs are a mix of Motorola MVME167, MVME177 and MVME2700 (PowerPC) models with either 16 or 32 MB of memory. The geographic layout of our accelerator led us to use most of these IOCs for multiple applications rather than follow the single application model used at many EPICS sites. Loading numerous applications on one IOC naturally taxes the computer resources more heavily than the single application per IOC model. Additionally, we have seen our IOC CPU loads grow as many of our applications have been updated to provide additional diagnostic information to the operators and new applications have been added to support more devices installed in the accelerator. These factors have left many IOCs in our system with woefully inadequate available CPU margins.

In an effort to provide enough headroom for comfortable operations, many of the accelerator IOCs have been upgraded from MVME167 or MVME177 processors to the more powerful MVME2700 Motorola PowerPC computers. This approach has provided the needed relief for our heaviest loaded IOCs, but it is an expensive and time-consuming process to test all the IOC applications on a new platform and make the code modifications that are necessary due to the drastically different architecture of the PowerPC. It is also very difficult to make such changes under the constraints of an operational machine with very limited scheduled maintenance time.

Due to the expense and difficulties involved in the processor upgrades, we also examined ways to reduce the IOC load in a more global fashion. Because name resolution is a task executed by each IOC every time a client looks for the location of a PV, and such searches are a very common occurrence, we examined this area. By default, EPICS clients searching for PVs broadcast to every IOC in the system. Each IOC then proceeds to determine if the requested PV is within its internal tables. The IOC that in fact hosts the requested PV then responds to the broadcasting client and the client establishes a connection with the IOC. The idea behind using a nameserver is to provide a central location where all clients can look for the address of the front-end computer with the desired PV and then make a direct IOC connection, thereby avoiding the load imposed on all system IOCs in the search by broadcast model.

Several years ago, Jefferson Lab implemented a Channel Access Nameserver[3] based on the Control Device API (CDEV) for use in our EPICS system, and this program was somewhat effective in reducing the IOC load due to name resolution. However, due to the difficulties associated with diverting client programs to query the CDEV Nameserver instead of broadcasting to the IOCs and the overhead needed to maintain the CDEV Nameserver through Channel Access upgrades, this program was used for only two clients, the Display Manager (MEDM) and the Save/Restore Tool (BURT). These two clients were selected because they accounted for the largest percentage of our name resolution traffic, however, it became clear our system would benefit from expanding this concept to all client programs due to growing demands on the IOCs. This led to the development of the New Channel Access Nameserver, detailed in this paper.

---
*Work supported by the U.S. Department of Energy under contract DE-AC05-84-ER40150

## 3 REQUIREMENTS

The primary function required of the new Nameserver was the ability to serve front-end computer IP addresses and port numbers to back-end clients. In order to avoid the weaknesses of the CDEV Nameserver, the new program was required to provide an easy mechanism for clients to switch between using nameservices for PV name resolution and the search by broadcast method. This requirement was placed to facilitate testing, and make the new Nameserver easy to use and maintainable. Another requirement was to provide a way for the new Nameserver to identify and learn the location of PVs added to the system after initialization, and update PV and address information following the reboot of an IOC.

## 4 IMPLEMENTATION

### 4.1 Operational Modes

The new Nameserver runs on a Unix workstation and is based on the EPICS Portable Channel Access Server (PCAS)[4] and the PCAS example program "directoryServer". Two modes of operation are available. In normal mode, information about the location of control system PVs is loaded into internal hashtables at startup. In learn mode, a request for information about an unknown PV triggers a broadcast by the Nameserver. If the requested PV is found, location information is added to the hashtable. At Jefferson Lab, we use a combination of both modes, preloading data for 98 IOCs and ~234k PVs and learning on demand about any other PVs. Preloading results in more deterministic operation and a startup time of less than 5 seconds.

### 4.2 Configuration

On bootup, each IOC creates a "signal.list" file using the command "dbl > signal.list". This file contains, in ASCII text, the name of every PV on the IOC. We load a standard system diagnostic EPICS database on each IOC and this database includes a PV of the form "iocname:heartbeat". At Jefferson Lab, these PVs are counters, updating every second and are used on MEDM screens to indicate that the IOC is running. The Nameserver utilizes a configuration file, with the defaultname "pvDirectory.txt", containing a list of full pathnames to the "signal.list" files for each control system IOC. The Nameserver reads the "pvDirectory.txt" and "signal.list" files during initial startup. If the file "pvDirectory" is empty, the Nameserver will run in total learn mode. This is not recommended for most efficient operation but may be useful for testing.

The requirement to provide a mechanism for easy switching between Nameserver and broadcast operations is accomplished through the use of environment variables. Normally, the EPICS environment variable, "EPICS_CA_ADDR_LIST" is set to the broadcast addresses of the control system IOCS. Any client can set its own "EPICS_CA_ADDR_LIST" to the IP address of the machine running the Nameserver, effectively preventing the normal broadcast to all control system IOCs and rerouting the name resolution request to only the Nameserver. In either case, the environment variable, "EPICS_CA_AUTO_ADDR_LIST" is set to "NO". This avoids problems with the clients seeing the PVs twice, once directly and again via the nameserver. This configuration mechanism allowed us to easily test and phase in the new Nameserver. After preliminary testing, all clients on specific console computers were set to divert name resolution to the Nameserver, and finally, all clients on all consoles were configured this way.

### 4.3 Software Description

The nameserver front end is modeled after the EPICS PCAS gateway[5] code and like the gateway, it may be run manually or as a daemon process. As a daemon process, the executable can be signaled to quit, to restart or to display run-time statistics. A log file may be created in manual mode and will be created if the server is run as a daemon.

The EPICS PCAS library provides many virtual functions, which are normally reimplemented by the application developer. The Nameserver uses only the "pvExistTest" function. The usual "read" and "write" functions are not reimplemented and effectively do nothing.

The library provides all low level code for channel access. On receipt of a client UDP message, the library code calls the application "pvExistTest" function with the PV name as an argument. In the nameserver code, the PV hashtable is checked. If the PV is found, the IOC hashtable is checked to verify that the IOC is up. If the IOC is up, the function returns "pvExistReturn" along with socket address information for the appropriate IOC. If the PV is not found in the PV hashtable, the Nameserver will attempt to learn the PV location by broadcasting to all control system IOCs and, if successful, will add that PV location to the hashtable.

The Nameserver maintains information on the status of each IOC by setting up a channel access monitor for each "heartbeat" PV and receiving channel access connection events. When a PV has been discovered rather than known via a "signal.list" file, the first PV requested and found on each previously unknown IOC is used as a channel access monitor for that IOC. In this way, a PV can be added, deleted or moved from one IOC to another and the disconnection/connection sequence from the IOC will trigger updates to both the status of the IOCs and the contents of the PV hashtable.

Our use of this centralized Nameserver also provides two useful features for managing a large and changing control system. We are able to accumulate a list of PVs, which the Nameserver is unable to resolve. This list has been helpful in cleaning up old code and configuration files, thus minimizing network traffic. In addition, duplicate PVs found during Nameserver initialization are immediately detected and logged.

The rest of the code is bookkeeping. In addition to the PCAS library, the Nameserver uses two EPICS hashtables to store information about PVs and IOCs and an EPICS singly linked list to keep track of PVs for which broadcasts have been done but have not yet been found or timed out. Unlike the earlier CDEV Nameserver, no modifications are necessary to either EPICS/base or to client code.

The only difficulty encountered with implementing the new Nameserver involves client code, which was compiled with very old versions of EPICS, using Channel Access protocol version 4.4 or older. In order to make these applications work with the new Nameserver they have been recompiled.

## 5 CONCLUSIONS

A comparison of the use of the new Nameserver, the CDEV based Nameserver and broadcast based name resolution shows that each method has advantages and disadvantages. EPICS broadcasts require no additional process or data management system. They do impose a load on the IOCs and network. The CDEV Nameserver lightens the IOC load but requires modification to EPICS channel access code and to any client, which uses the nameserver. These modifications must be done for each EPICS release and client. Requests for PVs unknown to this Nameserver "fall through" to EPICS broadcasts.

The new Nameserver currently in use requires no changes to EPICS or to client code, as redirection is simply handled by changing an environment variable. If a real PV unknown to the Nameserver is requested, the Nameserver is able to learn the location of the PV and update its internal tables. IOC reboots and corresponding changes in PV databases are handled automatically. The new Nameserver runs as a daemon process programmed to autorestart in the event of failure. Although we have not experience any failures of the new Nameserver, we have forced failures in order to test the system. In case of a failure, currently connected clients are unaffected and new connections are delayed for less than five seconds during restart and reloading of the 240K PV database. Timing tests comparing the connection speeds of clients using the new Nameserver and clients using broadcasts for PV name resolution showed that connections via the Nameserver were, on average, twice as fast.

## 7 ACKNOWLEDGEMENTS

Thanks to Jeff Hill for his suggestion of using directoryServer as a basis for the Nameserver and to the many people who contributed to the CA Gateway on which the Nameserver front-end is patterned.